\documentclass[12pt]{article}
\usepackage{amsfonts}
\usepackage{amsmath}
\usepackage{amssymb}
\usepackage{myart}

\normalbaselines
\font\tenbm=cmmib10
\font\sevenbm=cmmib7

\newfam\bmfam
\textfont\bmfam=\tenbm \scriptfont\bmfam=\sevenbm
\input tcilatex.tex
\begin{document}

\author{Yuri A. Rylov}
\title{Why does the Standard Model fail to explain the elementary particles
structure?}
\date{Institute for Problems in Mechanics, Russian Academy of Sciences \\
101-1 ,Vernadskii Ave., Moscow, 117526, Russia \\
email: rylov@ipmnet.ru\\
Web site: {$http://rsfq1.physics.sunysb.edu/\symbol{126}rylov/yrylov.htm$}\\
or mirror Web site: {$http://gasdyn-ipm.ipmnet.ru/\symbol{126}%
rylov/yrylov.htm$}}
\maketitle

\begin{abstract}
It is shown, that our contemporary knowledge of geometry is insufficient,
because we know only axiomatizable geometries. With such a knowledge of
geometry one cannot investigate properly physics of microcosm and structure
of elementary particles. One can obtain only a phenomenological systematics
of elementary particles, whose construction does not need a discrimination
mechanism. The discrimination mechanism, responsible for discrete
characteristics of elementary particles, can be created only on the basis of
a granular (discrete and continuous simultaneously) space-time geometry.
\end{abstract}

\section{Introduction}

Many theorists, dealing with the microcosm physics and with the theory of
elementary particles, believe, that the predictions of the Standard Model
will be confirmed by experiments on the large hadron collider. If so, we
shall understand the elementary particles arrangement.

I cannot agree with such an optimistic viewpoint. I think, that the Standard
Model, as well as other conceptions of the elementary particle theory
describe only systematization of elementary particles, but not their
arrangement.

Let me illustrate my statement in the example of the atomic theory. The atom
arrangement is described by means of the quantum mechanics, whereas the
systematization of chemical elements is described by the periodic system of
chemical elements. The theory of the atom arrangement and the periodic table
of chemical elements are quite different conceptions. The Standard Model as
well as other conceptions of the elementary particles theory suggest only
different methods of the elementary particles systematization, but not
different versions of the elementary particles arrangement. In this sense
the Standard Model is only an analog of the periodical system of chemical
elements, but not that of the atom arrangement theory.

The periodic system of the chemical elements was suggested by D.I. Mendeleev
in 1870. Mendeleev does not motivate his suggestion of the periodic system.
He said, that he had seen this system in a dream. The system predicted new
unknown elements and their properties. These elements were discovered, and
the trueness of the periodic system had been proved.

Fifty years later physicists began investigations of the atomic structure.
They used new quantum principles and succeeded in investigations of the
atoms arrangement.

Now, when we know the history, we may put the following question. Did the
periodic system help us in construction of the atomic theory. The answer is
negative. The two conceptions were developed by different investigators,
which had different education and tended to different goals. They used
different mathematical tools, and mathematical tools of physicists,
constructing the atomic theory, were more developed. In particular,
mathematical tools of physicists contained some discrimination mechanism,
which admits one to separate discrete values from the continual set of
possible values. From the mathematical viewpoint this discrimination
mechanism is the formalism of linear operators and their eigenvalues. From
physical viewpoint this mechanism is conditioned by the stabilizing role of
the atom electromagnetic emanation, which removes all nonstationary states,
remaining only stationary ones. Chemists, inventing and using the periodical
system had not such a discrimination mechanism, and they did not contribute
in the theory of atomic arrangement. As a result the periodic system of
chemical elements did not play any role in the construction of the atomic
theory. The Standard Model and other conceptions of the contemporary
elementary particle theory have no discrimination mechanism among their
mathematical tools, and they will not play any role in the construction of
the future theory of the elementary particle arrangement. This circumstance
does not exclude, that the Standard Model may be very useful for practical
investigations of the elementary particles properties, as well as the
periodic system of chemical elements is useful for practical work of
investigators-chemists.

Investigations of hadrons had lead to idea, that the hadrons have a
composite structure. It is supposed that hadrons consist of more elementary
particles, known as quarks. Attempts of extracting quarks from hadrons
failed. This phenomenon is known as confinement. Now there is no reasonable
explanation of the confinement phenomenon. The most simple and reasonable
explanation would be a reference to the properties of the space-time
geometry. If one supposes, that the space-time is discrete and the
geometrical objects cannot be divided into parts without limit, the
confinement may be easily explained by that circumstance, that hadrons are
"atoms of the space-time". They have composite structure. Nevertheless they
cannot be divided into parts. Besides, the supposition, that the space-time
geometry is discrete, admits one to understand the discrimination mechanism,
generated by the space-time geometry.

Let me note, that the supposition on discreteness of the space-time geometry
is not a hypothesis. In reality, the supposition, that any geometry (and
that of the space-time) is continual and divisible without limit, is a
hypothesis. This hypothesis was introduced in that time, when investigators
dealed with relatively large bodies. Their sizes were more large, than
possible elementary lengths, connected with the space-time discreteness.
Then it was unessential, whether the space-time is discrete or not. It was
unessential, whether the space-time geometry is divisible without limit or
not.

The problem of discreteness and of restricted divisibility of geometry is
not mentioned practically in the contemporary geometry. It is supposed, that
the geometry is continual and divisible without limit. Other versions of the
space-time geometry are not considered at all. It is connected with the
circumstance, that we are not able to work with such geometries. Our
knowledge of geometry is far from completeness.

In general, our approach to the space-time geometry must be as follows. We
do not adduce any suppositions on properties of the space-time geometry. We
must develop a space-time geometry of a general form. The real properties of
the space-time geometry must be determined from investigation of the real
bodies dynamics. It was made for distances approximately in the range $%
10^{-8}\div 10^{14}$cm. One has microcosm for distances less, than atomic
size $10^{-8}$cm, where the space-time geometry is not investigated
properly. One has megacosm for the distances larger, than the size of the
Solar system approximately $10^{14}$cm , where the space-time is not
investigated properly. Thus, the problem lies in our imperfect knowledge of
a geometry.

What is a geometry? The geometry is a science on mutual disposition of
geometrical objects in the space, or in the space-time. Geometry is a
continual set $\mathcal{S}$ of all propositions on properties of geometrical
objects. A geometry $\mathcal{G}_{\mathrm{a}}$ is axiomatizable, if the set $%
\mathcal{S}$ of all propositions can be deduced from the finite set $%
\mathcal{A}$ of basic propositions by means of the rules of the formal
logic. These basic propositions are known as axioms. The set $\mathcal{A}$
of axioms is called axiomatics of the given axiomatizable geometry $\mathcal{%
G}_{\mathrm{a}}$.

If one asks some person, having a humanitarian education, what is a
geometry, the answer will look something like that: "Geometry!? I studied it
in the school. It is something, when one proves different theorems and other
like things". If one puts the same question to professional
geometer-topologist, his answer will be very scientifically founded, but it
will distinguish from the answer of humanitarian educated person only in
some details. He will say: "Geometry is a set of propositions which is
deduced from axiomatics of the geometry." He will not mentioned, that the
geometry is axiomatizable, because, he knows only axiomatizable geometries,
and a mention on nonaxiomatizable geometries seems to him needless.

There is a paradoxical theorem of G\"{o}del, which may be formulated in the
form: "If we suppose that the geometry can be axiomatized, then it appears,
that the geometry cannot be axiomatized". Of course, it is a free paraphrase
of the G\"{o}del theorem. Nevertheless this theorem shows that a supposition
on possibility of a geometry axiomatization leads to paradoxical result.
This result means, that there exist nonaxiomatizable geometries.

However, what is a nonaxiomatizable geometry? It is the continual set $%
\mathcal{S}$ of all propositions on properties of geometrical objects, which
cannot be deduced from an axiomatics. In a sense, all propositions (or
continual set of them) are basic propositions, which cannot be deduced from
the axiomatics. In general, one cannot contradict anything against existence
of nonaxiomatizable geometries. However, how can one construct the continual
set of propositions, if one cannot use the formal logic for multiplication
of the geometrical propositions? The intuitively evident statement, that a
geometry (as a science on the mutual disposition of geometrical objects) is
determined completely, if the distance between any pair of points is given,
does not permit one to construct the continual set of all geometrical
propositions. Introduction of metric space, based on the idea of distance,
was not able to overcome the problem of construction of geometrical objects
and geometrical propositions in the metric space. As a result the metric
space does not generate a metric geometry.

There is a well known method of the geometry construction. If one deforms
the proper Euclidean geometry, i.e. if the distance between the points of
the Euclidean geometry is changed, one obtains another geometry, for
instance, the Riemannian geometry. This method of the geometry construction
does not refer to the axiomatizability of a geometry. It needs only, that
the geometry be described completely by the distance function between any
two points of the geometry. It is more convenient to use half of the squared
distance instead of the distance, because this quantity, known as the world
function, is real even in geometries with indefinite metric, for instance,
in the geometry of Minkowski. The geometry, which is described completely by
its world function is called the physical geometry, because such a geometry
is adequate for description of the space-time. The circumstance, whether the
physical geometry is axiomatizable or not, is not important for physicists.
It is important only, when the method of a geometry construction is founded
on the geometry axiomatization. One obtains a physical geometry as a
deformation of some standard geometry, which is axiomatizable and physical
simultaneously.

The proper Euclidean geometry $\mathcal{G}_{\mathrm{E}}$ may be used as such
a standard geometry, because the proper Euclidean geometry is axiomatizable 
\cite{H30} and physical \cite{R2001} simultaneously. As far as the proper
Euclidean geometry $\mathcal{G}_{\mathrm{E}}$ is an axiomatizable, all
propositions $\mathcal{S}_{\mathrm{E}}$ of $\mathcal{G}_{\mathrm{E}}$ can be
deduced from the axiomatics of $\mathcal{G}_{\mathrm{E}}$. As far as $%
\mathcal{G}_{\mathrm{E}}$ is a physical geometry, the continual set $%
\mathcal{S}_{\mathrm{E}}$ of all propositions of the standard geometry $%
\mathcal{G}_{\mathrm{E}}$ can be expressed in terms of the world function $%
\sigma _{\mathrm{E}}$ of the standard geometry $\mathcal{G}_{\mathrm{E}}$ in
the form $\mathcal{S}_{\mathrm{E}}=\mathcal{S}_{\mathrm{E}}\left( \sigma _{%
\mathrm{E}}\right) $. Deformation of the standard geometry $\mathcal{G}_{%
\mathrm{E}}$ means a replacement of the world function $\sigma _{\mathrm{E}}$
with some world function $\sigma $ of some other physical geometry $\mathcal{%
G}$. As a result of the deformation (replacement $\sigma _{\mathrm{E}%
}\rightarrow \sigma $), one obtains the set of all propositions $\mathcal{S}%
_{\mathrm{E}}\left( \sigma \right) $ of the physical geometry $\mathcal{G}$.

The deformation of an axiomatizable geometry $\mathcal{G}_{\mathrm{E}}$
transforms this geometry in a physical geometry $\mathcal{G}$, which is
nonaxiomatizable, in general, and the deformation method is a method of
nonaxiomatizable physical geometries construction. The axiomatizability of a
geometry is important only from the point of view of the geometry
construction. If one can construct nonaxiomatizable geometries, it is of no
importance, whether or not the geometry is axiomatizable. On the other hand,
a physical geometry possesses such properties, which cannot have the
axiomatizable geometry. The most interesting feature of a physical geometry
is that, the physical geometry may generate a discrimination mechanism,
which leads to discrete characteristics of particles, if the space-time is,
described by the proper nonaxiomatizable space-time geometry.

To understand this, let us consider the conventional (Euclidean) method of
the axiomatizable geometry construction. According to this method one needs
to postulate a system of axioms instead of the Euclidean axioms. Having
postulated a system of axioms, one needs to test compatibility of these
axioms between themselves. Compatibility of axioms means, that any
proposition of the geometry does not depend on the method of its deduction.
In practice, it means, that one needs to construct the continual set of all
propositions of the geometry and to test that different methods of the
deduction of a proposition lead to the same result. Of course, it is very
difficult task, and nobody test compatibility of all axioms of the geometry.
Instead of the test everybody believe, that the axiomatics of the geometry
is consistent, and construct those propositions, which are interesting in
the considered problem.

The problem of the physical geometry consistency, constructed by the
deformation method, is absent at all, because it is a problem of the method
of the geometry construction, but not the problem of the geometry in itself.
This is the first advantage of the deformation method. To obtain some
proposition of the geometry by means of the Euclidean method, one needs to
formulate some theorem and prove it. In many cases the procedure of the
proof appears to be rather complicated. Using the deformation method, one
does not need to prove any theorems, to obtain any proposition of the
physical geometry. The proposition of the physical geometry $\mathcal{G}$ is
obtained from the standard geometry $\mathcal{G}_{\mathrm{E}}$ after
replacement of the world function $\sigma _{\mathrm{E}}$ by the world
function $\sigma $ in the corresponding proposition of the standard geometry.

The physical geometry is formulated in terms of points and world functions
between these points. At formulation of the physical geometry propositions
one does not use such non-invariant methods of description, which refer to
manifold, coordinate system and dimension. The proper Euclidean (standard)
geometry is given as a rule on a manifold in some coordinate system. To
deform the proper Euclidean geometry, one needs to represent it in the $%
\sigma $-immanent form, i.e. in the form, which does not refer to coordinate
system and contains only points and world functions between them. In some
cases such a transformation of the conventional description (in the
coordinate form) to the $\sigma $-immanent representation may be rather
difficult and unexpected. But these problems are problems of the proper
Euclidean geometry $\mathcal{G}_{\mathrm{E}}$, and they can be solved,
provided we know the proper Euclidean geometry well enough. Any proposition
of the Euclidean geometry $\mathcal{G}_{\mathrm{E}}$ can be expressed in the 
$\sigma $-immanent form always. There is a theorem on that score \cite{R2001}%
.

As a rule a physical geometry is nonaxiomatizable and has very important
properties, which are new for axiomatizable geometries. The general name for
these properties is multivariance. To obtain these properties, let us
consider the property of equivalence of two vectors $\mathbf{P}_{0}\mathbf{P}%
_{1}$ and $\mathbf{Q}_{0}\mathbf{Q}_{1}$ in the proper Euclidean geometry $%
\mathcal{G}_{\mathrm{E}}$. The geometry is given on the point set $\Omega $.
Vector $\mathbf{P}_{0}\mathbf{P}_{1}\equiv \overrightarrow{P_{0}P_{1}}%
=\left\{ P_{0},P_{1}\right\} $ is an ordered set of two points $P_{0}$ and $%
P_{1}$. The length $\left\vert \mathbf{P}_{0}\mathbf{P}_{1}\right\vert $ of
the vector $\mathbf{P}_{0}\mathbf{P}_{1}$ is defined by the relation%
\begin{equation}
\left\vert \mathbf{P}_{0}\mathbf{P}_{1}\right\vert =\sqrt{\left( \mathbf{P}%
_{0}\mathbf{P}_{1}.\mathbf{P}_{0}\mathbf{P}_{1}\right) }=\sqrt{2\sigma
\left( P_{0},P_{1}\right) }  \label{a1.1}
\end{equation}%
where $\sigma \left( P_{0},P_{1}\right) $ is the world function 
\begin{equation}
\sigma :\qquad \Omega \times \Omega \rightarrow \mathbb{R},\qquad \sigma
\left( P,P\right) =0,\qquad \forall P\in \Omega  \label{a1.2}
\end{equation}

The scalar product $\left( \mathbf{P}_{0}\mathbf{P}_{1}.\mathbf{Q}_{0}%
\mathbf{Q}_{1}\right) $ of two vectors $\mathbf{P}_{0}\mathbf{P}_{1}$ and $%
\mathbf{Q}_{0}\mathbf{Q}_{1}$ is defined by the relation%
\begin{equation}
\left( \mathbf{P}_{0}\mathbf{P}_{1}.\mathbf{Q}_{0}\mathbf{Q}_{1}\right)
=\sigma \left( P_{0},Q_{1}\right) +\sigma \left( P_{1},Q_{0}\right) -\sigma
\left( P_{0},Q_{0}\right) -\sigma \left( P_{1},Q_{1}\right)  \label{a1.3}
\end{equation}%
In the given case the relations (\ref{a1.1}) -- (\ref{a1.3}) are written for
the proper Euclidean geometry $\mathcal{G}_{\mathrm{E}}$, and $\sigma
=\sigma _{\mathrm{E}}$ is the world function of $\mathcal{G}_{\mathrm{E}}$.
However, these relations are valid in any physical geometry. In $\mathcal{G}%
_{\mathrm{E}}$ one can easily verify, that the definition of the scalar
product (\ref{a1.3}) coincides with the conventional definition of the
scalar product. In $\mathcal{G}_{\mathrm{E}}$ two vectors $\mathbf{P}_{0}%
\mathbf{P}_{1}$ and $\mathbf{Q}_{0}\mathbf{Q}_{1}$ are equivalent (equal) $%
\mathbf{P}_{0}\mathbf{P}_{1}$eqv$\mathbf{Q}_{0}\mathbf{Q}_{1}$, if 
\begin{equation}
\mathbf{P}_{0}\mathbf{P}_{1}\mathrm{eqv}\mathbf{Q}_{0}\mathbf{Q}_{1}:\qquad
\left( \mathbf{P}_{0}\mathbf{P}_{1}.\mathbf{Q}_{0}\mathbf{Q}_{1}\right)
=\left\vert \mathbf{P}_{0}\mathbf{P}_{1}\right\vert \cdot \left\vert \mathbf{%
Q}_{0}\mathbf{Q}_{1}\right\vert \wedge \left\vert \mathbf{P}_{0}\mathbf{P}%
_{1}\right\vert =\left\vert \mathbf{Q}_{0}\mathbf{Q}_{1}\right\vert
\label{a1.4}
\end{equation}%
The same definition (\ref{a1.4}) is true in any physical geometry.

The definition (\ref{a1.4}) means that in any physical geometry there is an
absolute parallelism, which is described by the first relation (\ref{a1.4}).
In the (pseudo-) Riemannian geometry, which is used usually as the
space-time geometry, there is no absolute parallelism, in general. Does it
mean, that the Riemannian geometry is not a physical geometry? Later on I
shall return to this interesting problem.

Let vector $\mathbf{Q}_{0}\mathbf{Q}_{1}$ be given at the point $Q_{0}$, and
one tries to determine an equivalent vector $\mathbf{P}_{0}\mathbf{P}_{1}$
at the point $P_{0}$. Let for simplicity the geometry is given on the
four-dimensional manifold $\Omega $. Coordinates of points $%
P_{0},Q_{0},Q_{1} $ are given. Four coordinates of the point $P_{1}$ are to
be determined as a solution of two equations (\ref{a1.4}) with the scalar
product $\left( \mathbf{P}_{0}\mathbf{P}_{1}.\mathbf{Q}_{0}\mathbf{Q}%
_{1}\right) $, given by the relation (\ref{a1.3}). In the proper Euclidean
geometry $\mathcal{G}_{\mathrm{E}}$ four coordinates of the point $P_{1}$
are determined by the two relations (\ref{a1.4}) single-valuedly, although
the number of coordinates is four, whereas the number of equations is two.
Such a single-valuedness is a corollary of special properties of $\mathcal{G}%
_{\mathrm{E}}$. It is valid for the Euclidean geometry $\mathcal{G}_{\mathrm{%
E}}$ of any dimension. If the geometry $\mathcal{G}$ is the geometry of
Minkowski, one obtains a unique solution for the timelike vector $\mathbf{Q}%
_{0}\mathbf{Q}_{1}$. If the vector $\mathbf{Q}_{0}\mathbf{Q}_{1}$ is
spacelike, the number of solution for the point $P_{1}$ is infinite. In
other words, at the point $P_{0}$ there are many vectors $\mathbf{P}_{0}%
\mathbf{P}_{1}$ ,$\mathbf{P}_{0}\mathbf{P}_{1}^{\prime }$, $\mathbf{P}_{0}%
\mathbf{P}_{1}^{\prime \prime }$,..., which are equivalent to the vector $%
\mathbf{Q}_{0}\mathbf{Q}_{1}$, but they are not equivalent between
themselves. Such a property of the physical geometry will be referred to as
multivariance of the geometry $\mathcal{G}$ with respect to the point $P_{0}$
and the vector $\mathbf{Q}_{0}\mathbf{Q}_{1}$. Multivariance of the geometry 
$\mathcal{G}$ is possible, only if the equivalence relation in the geometry $%
\mathcal{G}$ is intransitive. In any axiomatizable geometry the equivalence
relation is always transitive.

Thus, an axiomatizable geometry cannot be multivariant. On the other hand,
the multivariance is a natural property of a physical geometry, because one
cannot guarantee existence and uniqueness of a solution of the equations (%
\ref{a1.4}) for arbitrary world function $\sigma $. As a rule the physical
geometries are multivariant, and the equivalence relation in them is
intransitive. It means, that the physical geometries are nonaxiomatizable as
a rule. On the other hand, in any axiomatizable geometry the equivalence
relation is transitive, and an axiomatizable geometry cannot be multivariant.

It is reasonable, that adherers of the conventional Euclidean method of the
geometry construction cannot imagine existence of nonaxiomatizable
geometries. If some geometry manifests some evidence of multivariance, it
means, that this geometry is nonaxiomatizable. From the viewpoint of
adherers of the Euclidean method the nonaxiomatizable geometries do not
exist. From their viewpoint the multivariance of a geometry means, that
axiomatics of this geometry is defective (maybe, inconsistent). To remove
defects from the axiomatics, one needs to remove multivariance from the
geometry.

The Riemannian geometry manifests evidence of multivariance. Using
conventional methods of the Riemannian geometry construction one cannot
define absolute parallelism in the Riemannian geometry. The world function $%
\sigma _{\mathrm{R}}$ of the Riemannian geometry $\mathcal{G}_{\mathrm{R}}$
is defined by the relation 
\begin{equation}
\sigma _{\mathrm{R}}\left( P_{0},P_{1}\right) =\frac{1}{2}\left(
\dint\limits_{\mathcal{L}_{P_{0}P_{1}}}\sqrt{g_{ik}\left( x\right)
dx^{i}dx^{k}}\right) ^{2}  \label{a1.5}
\end{equation}%
where the integral is taken along the geodesic $\mathcal{L}_{P_{0}P_{1}}$,
connecting points $P_{0}$ and $P_{1}$.

Taking the world function (\ref{a1.5}) as the world function $\sigma
_{\sigma \mathrm{R}}$ of a physical geometry $\mathcal{G}_{\sigma \mathrm{R}%
} $, and constructing the physical geometry $\mathcal{G}_{\sigma \mathrm{R}}$%
, one discovers that $\mathcal{G}_{\sigma \mathrm{R}}$ is multivariant. In
particular, the straight (geodesic) $\mathcal{L}_{Q_{0};\mathbf{P}_{0}%
\mathbf{P}_{1}}$, passing through the point $Q_{0}$ in parallel with vector $%
\mathbf{P}_{0}\mathbf{P}_{1}$, is a hallow tube, but not a one-dimensional
line. In the case, when the point $Q_{0}$ coincides with the point $P_{0}$
(or $P_{1}$) the straight (geodesic) $\mathcal{L}_{P_{0};\mathbf{P}_{0}%
\mathbf{P}_{1}}$ degenerates into one-dimensional line. If the Riemannian
geometry is an axiomatizable geometry, it cannot be multivariant. To
eliminate the multivariance, one declares, that the absolute parallelism is
absent in the Riemannian geometry, and one cannot construct the geodesic $%
\mathcal{L}_{Q_{0};\mathbf{P}_{0}\mathbf{P}_{1}}$ with $Q_{0}\neq P_{0}$.
The geometry $\mathcal{G}_{\sigma \mathrm{R}}$ is nonaxiomatizable. Imposing
an additional constraint, can one be sure, that this constraint makes the
geometry axiomatizable? Of course, no, because the multivariance may appear
in other propositions of the geometry.

Strictly, if one believes, that some geometry is axiomatizable and
consistent, one needs to prove these statements. One needs to formulate
axiomatics and prove its consistency. As far as I know, nobody had proved
consistency of the Riemannian geometry. On the other side, the physical
geometry $\mathcal{G}_{\sigma \mathrm{R}}$ is nonaxiomatizable. There is no
question about its consistency, because this question relates to the
Euclidean method of a geometry construction. Imposing additional constraints
on the physical geometry, one cannot be sure, that the physical geometry
with additional constraint is a true geometry.

Besides, why does one think, that the multivariance is an alien property of
the geometry? It is true, that the multivariance is alien to axiomatizable
geometries, constructed by the Euclidean method. In reality, appearance of
multivariance in the Riemannian geometry, which may be used as a space-time
geometry, means that the multivariant nonaxiomatizable space-time geometries
exist, and one has no reason to ignore them.

If one investigates the problem, what is the geometry of the real
space-time, one should consider the most general geometries, including
multivariant nonaxiomatizable ones. After investigation of properties of all
possible space-time geometries and particle dynamics in them, one could
decide, which of these possible space-time geometries is realized in the
real space-time. The approach, when one discriminates nonaxiomatizable
geometries, is a preconceived approach, which shows, that our knowledge of
geometry is insufficient. In particular, choosing between two space-time
geometries: the Riemannian geometry and the physical geometry $\mathcal{G}%
_{\sigma \mathrm{R}}$, having the same world function, one should prefer the
geometry $\mathcal{G}_{\sigma \mathrm{R}}$, because at construction of the
Riemannian geometry one uses many amotivational constraints (continuity,
unlimited divisibility, use of manifold), which are absent at construction
of the physical geometry $\mathcal{G}_{\sigma \mathrm{R}}$. Besides, the
conventional Riemannian geometry may appear to be inconsistent, because its
consistency has not yet been proved. For the physical geometry $\mathcal{G}%
_{\sigma \mathrm{R}}$ the problem of inconsistency is absent at all.

\section{Unaccustomed properties of physical geometries}

I shall try to manifest unaccustomed and unexpected properties of a physical
space-time geometry in the example of the geometry $\mathcal{G}_{\mathrm{g}}$%
, described by the world function%
\begin{equation}
\sigma _{\mathrm{g}}=\sigma _{\mathrm{M}}+\lambda _{0}^{2}\left\{ 
\begin{array}{lll}
\mathrm{sgn}\left( \sigma _{\mathrm{M}}\right) & \text{if} & \left\vert
\sigma _{\mathrm{M}}\right\vert >\sigma _{0} \\ 
\frac{\sigma _{\mathrm{M}}}{\sigma _{0}} & \text{if} & \left\vert \sigma _{%
\mathrm{M}}\right\vert \leq \sigma _{0}%
\end{array}%
\right. ,\qquad \lambda _{0}^{2},\sigma _{0}=\text{const}\geq 0  \label{a2.1}
\end{equation}%
\begin{equation}
\text{sgn}\left( x\right) =\left\{ 
\begin{array}{lll}
\frac{x}{\left\vert x\right\vert } & \text{if} & x\neq 0 \\ 
0 & \text{if} & x=0%
\end{array}%
\right.  \label{a2.1a}
\end{equation}%
where $\sigma _{\mathrm{M}}$ is the world function of the geometry of
Minkowski. In the inertial coordinate system it has the form 
\begin{equation}
\sigma _{\mathrm{M}}\left( x,x^{\prime }\right) =\frac{1}{2}g_{ik}\left(
x^{i}-x^{\prime i}\right) \left( x^{k}-x^{\prime k}\right) ,\qquad g_{ik}=%
\text{diag}\left( c^{2},-1,-1,-1\right)  \label{a2.2}
\end{equation}%
Here $\lambda _{0}$ is some elementary length and $c$ is the speed of the
light. The geometry $\mathcal{G}_{\mathrm{g}}$ is given on the 4-dimensional
manifold, but this geometry is not continuous. The world function $\mathcal{%
\sigma }_{\mathrm{g}}$ is Lorentz-invariant, because $\sigma _{\mathrm{g}}$
is a function of $\sigma _{\mathrm{M}}$, and $\sigma _{\mathrm{M}}$ is
Lorentz-invariant. The elementary length $\lambda _{0}$ is a small quantity,
and $\sigma _{\mathrm{g}}\approx \sigma _{\mathrm{M}}$, if characteristic
sizes of the problem are much larger, than $\lambda _{0}$. In the microcosm,
where the characteristic lengths are of the order of $\lambda _{0}$, the
world functions $\sigma _{\mathrm{g}}$ and $\sigma _{\mathrm{M}}$
distinguish essentially.

As it is follows from (\ref{a2.1}), the relative density $\rho \left( \sigma
_{\mathrm{g}}\right) $ of points in the geometry $\mathcal{G}_{\mathrm{g}}$
with respect to the geometry of Minkowski $\mathcal{G}_{\mathrm{M}}$ is
described by the relation%
\begin{equation}
\rho \left( \sigma _{\mathrm{g}}\right) =\frac{d\sigma _{\mathrm{M}}\left(
\sigma _{\mathrm{g}}\right) }{d\sigma _{\mathrm{g}}}=\left\{ 
\begin{array}{lll}
1 & \text{if} & \left\vert \sigma _{\mathrm{g}}\right\vert >\sigma
_{0}+\lambda _{0}^{2} \\ 
\frac{\sigma _{0}}{\sigma _{0}+\lambda _{0}^{2}} & \text{if} & \left\vert
\sigma _{\mathrm{g}}\right\vert \leq \sigma _{0}+\lambda _{0}^{2}%
\end{array}%
\right.  \label{a2.3}
\end{equation}%
If $\sigma _{0}\rightarrow 0$, the geometry $\mathcal{G}_{\mathrm{g}}$ tends
to the geometry $\mathcal{G}_{\mathrm{d}}$, described by the world function $%
\sigma _{\mathrm{d}}$%
\begin{equation}
\sigma _{\mathrm{d}}=\sigma _{\mathrm{M}}+d\mathrm{sgn}\left( \sigma _{%
\mathrm{M}}\right) ,\qquad d\equiv \lambda _{0}^{2}=\text{const}
\label{a2.3a}
\end{equation}%
The relative density $\rho \left( \sigma _{\mathrm{d}}\right) $ of points in
the geometry $\mathcal{G}_{\mathrm{d}}$ with respect to the geometry of
Minkowski $\mathcal{G}_{\mathrm{M}}$ is described by the relation 
\begin{equation}
\rho \left( \sigma _{\mathrm{d}}\right) =\frac{d\sigma _{\mathrm{M}}\left(
\sigma _{\mathrm{d}}\right) }{d\sigma _{\mathrm{d}}}=\left\{ 
\begin{array}{lll}
1 & \text{if} & \left\vert \sigma _{\mathrm{g}}\right\vert >\lambda _{0}^{2}
\\ 
0 & \text{if} & \left\vert \sigma _{\mathrm{g}}\right\vert \leq \lambda
_{0}^{2}%
\end{array}%
\right.  \label{a2.4}
\end{equation}%
As it is follows from (\ref{a2.4}) in the geometry $\mathcal{G}_{\mathrm{d}}$
there no close points, i.e. such points that the distance between them be
less, than the elementary length $\lambda _{0}/\sqrt{2}$. It means, that the
geometry $\mathcal{G}_{\mathrm{d}}$ is a discrete geometry. The geometry $%
\mathcal{G}_{\mathrm{d}}$ is a discrete geometry, although it is given on a
continuous manifold. It seems to be rather unexpected, that a discrete
geometry may be given on a manifold. It means that the physical geometry is
determined only by the form of its world function, but not by a character of
the point set, where the geometry is given.

Besides, one can imagine such a physical geometry, which is intermediate
between the continuous geometry and the discrete one. For instance, the
physical geometry $\mathcal{G}_{\mathrm{g}}$ is partly continuous geometry
and partly discrete geometry, because the point density $0<\rho \left(
\sigma _{\mathrm{g}}\right) <1$ in the region $\left\vert \sigma _{\mathrm{d}%
}\right\vert \leq \sigma _{0}+\lambda _{0}^{2}$ ( for discrete geometry $%
\rho \left( \sigma _{\mathrm{d}}\right) =0$, and for continuous geometry $%
\rho \left( \sigma _{\mathrm{d}}\right) =1$). I shall refer to such a
geometry $\mathcal{G}_{\mathrm{g}}$ as a granular geometry. This geometry $%
\mathcal{G}_{\mathrm{g}}$ turns into a discrete geometry $\mathcal{G}_{%
\mathrm{d}}$, if the constant $\sigma _{0}\rightarrow 0$. It turns into a
continuous geometry $\mathcal{G}_{\mathrm{M}}$, if $\lambda _{0}\rightarrow
0 $.

The granular space-time geometry distinguishes from the Riemannian
space-time geometry in the relation, that the granular geometry admits one
to formulate the particle dynamics in geometrical terms (points and world
function), i.e. without a reference to the coordinate system and
differential dynamic equations. Any (composite) particle is described by its
skeleton $\mathcal{P}_{n}=\left\{ P_{0},P_{1},..,P_{n}\right\} \in \Omega
^{n+1}$, where, $\Omega $ is the event set of the space-time. Evolution of
the particle skeleton $\mathcal{P}_{n}$ is described by the world chain of
skeletons ...$\mathcal{P}_{n}^{\left( 1\right) },\mathcal{P}_{n}^{\left(
2\right) }$...$\mathcal{P}_{n}^{\left( s\right) },$... Direction of
evolution is described by the leading vector $\mathbf{P}_{0}^{\left(
s\right) }\mathbf{P}_{1}^{\left( s\right) }$ in the sense, that it is
supposed that 
\begin{equation}
P_{0}^{\left( s+1\right) }=P_{1}^{\left( s\right) },\qquad s=...0,1,...
\label{a2.5}
\end{equation}%
If the particle is free, one has for links of the world chain%
\begin{equation}
\mathcal{P}_{n}^{\left( s+1\right) }\mathrm{eqv}\mathcal{P}_{n}^{\left(
s\right) },\qquad s=...0,1,...  \label{a2.6}
\end{equation}%
In the developed form the equations (\ref{a2.6}) mean%
\begin{equation}
\mathbf{P}_{k}^{\left( s+1\right) }\mathbf{P}_{l}^{\left( s+1\right) }%
\mathrm{eqv}\mathbf{P}_{k}^{\left( s\right) }\mathbf{P}_{l}^{\left( s\right)
},\qquad k<l,\qquad k,l=0,1,...n,\qquad s=...0,1,...  \label{a2.7}
\end{equation}%
The equation (\ref{a1.4}) can be represented in the form, which is linear
with respect to the world function%
\begin{equation}
\mathbf{P}_{0}\mathbf{P}_{1}\mathrm{eqv}\mathbf{Q}_{0}\mathbf{Q}_{1}:\qquad
\left( \mathbf{P}_{0}\mathbf{P}_{1}.\mathbf{Q}_{0}\mathbf{Q}_{1}\right)
=\left\vert \mathbf{P}_{0}\mathbf{P}_{1}\right\vert ^{2}\wedge \left\vert 
\mathbf{P}_{0}\mathbf{P}_{1}\right\vert ^{2}=\left\vert \mathbf{Q}_{0}%
\mathbf{Q}_{1}\right\vert ^{2}  \label{a2.7a}
\end{equation}%
Then equations (\ref{a2.7}) are written in the form%
\begin{eqnarray}
\left( \mathbf{P}_{k}^{\left( s+1\right) }\mathbf{P}_{l}^{\left( s+1\right)
}.\mathbf{P}_{k}^{\left( s\right) }\mathbf{P}_{l}^{\left( s\right) }\right)
&=&\left\vert \mathbf{P}_{k}^{\left( s\right) }\mathbf{P}_{l}^{\left(
s\right) }\right\vert ^{2},\quad k,l=0,1,...n,\quad s=...0,1,..  \label{a2.8}
\\
\left\vert \mathbf{P}_{k}^{\left( s+1\right) }\mathbf{P}_{l}^{\left(
s+1\right) }\right\vert ^{2} &=&\left\vert \mathbf{P}_{k}^{\left( s\right) }%
\mathbf{P}_{l}^{\left( s\right) }\right\vert ^{2},\quad k,l=0,1,...n,\quad
s=...0,1,..  \label{a2.9}
\end{eqnarray}

The free motion of a composite particle, described in the granular
space-time geometry $\mathcal{G}_{\mathrm{g}}$, can be described as a motion
in the some force field in the Kaluza-Klein geometry $\mathcal{G}_{\mathrm{K}%
}$. This transition reminds the case, when the free particle motion in the
Riemannian space-time geometry is substituted by the particle motion in the
gravitational field, given in the space-time of Minkowski.

To realize description of a composite particle in the Kaluza-Klein geometry,
one represents the world function $\sigma _{\mathrm{g}}$ of the granular
space-time geometry $\mathcal{G}_{\mathrm{g}}$ in the form%
\begin{equation}
\sigma _{\mathrm{g}}\left( P,Q\right) =\sigma _{\mathrm{K}}\left( P,Q\right)
+D\left( P,Q\right) ,\qquad \forall P,Q\in \Omega  \label{a2.10}
\end{equation}%
where $\sigma _{\mathrm{K}}$ is the world function of the Kaluza-Klein
geometry $\mathcal{G}_{\mathrm{K}}$. The geometry $\mathcal{G}_{\mathrm{K}}$
includes description of classical (gravitational and electromagnetic)
fields, and $D\left( P,Q\right) $ is the difference between the true world
function $\sigma _{\mathrm{g}}$ and the world function $\sigma _{\mathrm{K}}$%
, taking into account only classical fields. Then we have 
\begin{equation}
\left( \mathbf{P}_{k}^{\left( s\right) }\mathbf{P}_{l}^{\left( s\right) }.%
\mathbf{P}_{k}^{\left( s+1\right) }\mathbf{P}_{l}^{\left( s+1\right)
}\right) _{\mathrm{g}}=\left( \mathbf{P}_{k}^{\left( s+1\right) }\mathbf{P}%
_{l}^{\left( s+1\right) }.\mathbf{P}_{k}^{\left( s\right) }\mathbf{P}%
_{l}^{\left( s\right) }\right) _{\mathrm{K}}+w\left( P_{k}^{\left( s\right)
},P_{l}^{\left( s\right) },P_{k}^{\left( s+1\right) },\mathbf{P}_{l}^{\left(
s+1\right) }\right)  \label{a2.11}
\end{equation}%
where indices "g" and "K" mean that the scalar products are calculated
respectively in the granular geometry and the Kaluza-Klein geometry. The
quantity $w\left( P_{0},P_{1},Q_{0},Q_{1}\right) $ has the form 
\begin{equation}
w\left( P_{0},P_{1},Q_{0},Q_{1}\right) =D\left( P_{0},Q_{1}\right) +D\left(
P_{1},Q_{0}\right) -D\left( P_{0},Q_{0}\right) -D\left( P_{1},Q_{1}\right)
\label{a2.12}
\end{equation}%
Dynamic equations (\ref{a2.8}), (\ref{a2.9}) may be rewritten in the form%
\begin{eqnarray}
\left( \mathbf{P}_{k}^{\left( s+1\right) }\mathbf{P}_{l}^{\left( s+1\right)
}.\mathbf{P}_{k}^{\left( s\right) }\mathbf{P}_{l}^{\left( s\right) }\right)
_{\mathrm{K}} &=&2\sigma _{\mathrm{K}}\left( P_{k}^{\left( s\right)
},P_{l}^{\left( s\right) }\right) +2D\left( P_{k}^{\left( s\right)
},P_{l}^{\left( s\right) }\right)  \notag \\
&&+w\left( P_{k}^{\left( s+1\right) },P_{l}^{\left( s+1\right)
},P_{k}^{\left( s\right) },P_{l}^{\left( s\right) }\right)  \label{a2.14}
\end{eqnarray}%
\begin{eqnarray}
\sigma _{\mathrm{K}}\left( P_{k}^{\left( s+1\right) },P_{l}^{\left(
s+1\right) }\right) &=&\sigma _{\mathrm{K}}\left( P_{k}^{\left( s\right)
},P_{l}^{\left( s\right) }\right) \mathbf{+}D\left( P_{k}^{\left( s\right)
},P_{l}^{\left( s\right) }\right)  \notag \\
-D\left( P_{k}^{\left( s+1\right) },P_{l}^{\left( s+1\right) }\right)
,\qquad k &<&l,\qquad k,l=0,1,...n,\qquad s=...,0,1,..  \label{a2.16}
\end{eqnarray}%
where%
\begin{eqnarray}
w\left( P_{k}^{\left( s+1\right) },P_{l}^{\left( s+1\right) },P_{k}^{\left(
s\right) },P_{l}^{\left( s\right) }\right) &=&D\left( P_{k}^{\left(
s+1\right) },P_{l}^{\left( s\right) }\right) +D\left( P_{l}^{\left(
s+1\right) },P_{k}^{\left( s\right) }\right)  \notag \\
&&-D\left( P_{k}^{\left( s+1\right) },P_{k}^{\left( s\right) }\right)
-D\left( P_{l}^{\left( s+1\right) },P_{l}^{\left( s\right) }\right)
\label{a2.17}
\end{eqnarray}

In equations (\ref{a2.14}) -- (\ref{a2.17}) the classical fields (the
electromagnetic field and the gravitational field) are included in the
space-time geometry. They are described by the world function $\sigma _{%
\mathrm{K}}$. The force fields, characteristic for microcosm, have been
included in the function $D$. However, one can include the classical fields
in the function $D$, describing the force fields of the microcosm. Then the
world function $\sigma _{\mathrm{K}}$ will be describe the Kaluza-Klein
space-time, which is free of classical fields.

Let us note that dynamics of a free composite particle is described in terms
of the world function and points. It does not contain a reference to a
coordinate system, to continuity, or to other special properties of the
space-time. Dynamic equations are written in any physical space-time
geometry.

If the manifold, where the space-time geometry is given has the dimension $%
n_{\mathrm{K}}$, and $n+1$ is the number of points of the skeleton $\mathcal{%
P}_{n}$, the number of equations is equal to $n\left( n+1\right) $, whereas
the number of coordinates to be determined is equal to $n_{\mathrm{K}}n$.
These numbers coincide, if $n=n_{\mathrm{K}}-1$. In this case one should
expect, that the dynamic equations have an unique solution. However, it is
valid only in the case, when the leading vector $\mathbf{P}_{0}\mathbf{P}%
_{1} $, determining the direction of the particle evolution, is timelike.

If the leading vector $\mathbf{P}_{0}\mathbf{P}_{1}$ is spacelike, the
skeleton world chain may exist only, if it is a spacelike helix with
timelike axis. This condition imposes additional constraints on the dynamic
equations.

Let us consider an example. The classical limit of the Dirac equation
describes the classical Dirac particle $\mathcal{S}_{\mathrm{Dcl}}$. World
line of the free classical Dirac particle is a helix. It is not quite clear,
whether the world line is timelike, or spacelike, because the classical
Dirac particle appears to be composite \cite{R2004}, and its internal
degrees of freedom are described nonrelativistically \cite{R2004a}. The axis
of the helix is timelike. Dynamic equations, describing the classical\ Dirac
particle, contain the quantum constant, but they do not contain $\gamma $%
-matrices, which are characteristic for description of the quantum Dirac
particle. In the paper \cite{R2008} one puts the following question. Is it
possible, that the geometric dynamics (\ref{a2.14}) -- (\ref{a2.17})
describe a composite particle with the spacelike leading vector $\mathbf{P}%
_{0}\mathbf{P}_{1}$? It appears, that it is impossible for the space-time
geometry, described by the world function (\ref{a2.1}). However, it is
possible for the space-time geometry with the world function 
\begin{equation}
\sigma =\sigma _{\mathrm{M}}+\lambda _{0}^{2}\left\{ 
\begin{array}{lll}
\mathrm{sgn}\left( \sigma _{\mathrm{M}}\right) & \text{if} & \left\vert
\sigma _{\mathrm{M}}\right\vert >\sigma _{0} \\ 
\left( \frac{\sigma _{\mathrm{M}}}{\sigma _{0}}\right) ^{3} & \text{if} & 
\left\vert \sigma _{\mathrm{M}}\right\vert \leq \sigma _{0}%
\end{array}%
\right. ,\qquad \lambda _{0}^{2},\sigma _{0}=\text{const}\geq 0
\label{a2.18}
\end{equation}%
In this case the world chain is a spacelike helix with a timelike axis. The
particle is composite in the sense, that the skeleton consist of not less,
than three points. Additional points are needed for stabilization of the
helical world chain. Besides, the parameters of the helix cannot be
arbitrary. The helical world chain is possible only for some discrete values
of parameters. The consideration was produced on the four-dimensional
manifold of Minkowski, i.e. for the Dirac particle of zeroth charge. To
approach to the real situation, the spacelike world chain should be
considered on the five-dimensional manifold of Kaluza-Klein. However, even
such a model consideration on the manifold of Minkowski has shown, that the
physical granular space-time geometry can generate some discrimination
mechanism, responsible for discrete values of the particle characteristics.

It is worth to remark, that in the Riemannian geometry the spacelike world
line of a particle is impossible in principle, and the phenomenon of the
classical Dirac particle cannot be understood. In the granular space-time
geometry with the world function%
\begin{equation}
\sigma =\sigma _{\mathrm{M}}+\lambda _{0}^{2}\left\{ 
\begin{array}{lll}
\mathrm{sgn}\left( \sigma _{\mathrm{M}}\right) & \text{if} & \left\vert
\sigma _{\mathrm{M}}\right\vert >\sigma _{0} \\ 
f\left( \frac{\sigma _{\mathrm{M}}}{\sigma _{0}}\right) & \text{if} & 
\left\vert \sigma _{\mathrm{M}}\right\vert \leq \sigma _{0}%
\end{array}%
\right. ,\qquad \lambda _{0}^{2},\sigma _{0}=\text{const}\geq 0
\label{a2.19}
\end{equation}%
\begin{equation}
f\left( x\right) =-f\left( -x\right) ,\qquad x\in \left[ -1,1\right] ,\qquad
\left\vert f\left( x\right) \right\vert <\left\vert x\right\vert
\label{a2.20}
\end{equation}%
the spacelike helical world chain is possible for some discrete parameters
of the composite particle. Thus, the granular space-time geometry may
generate a discrimination mechanism.

The discrimination mechanism can be generated also by the space-time
compactification, which is essential for the Kaluza-Klein space-time
geometry \cite{R2008a}.

The granular space-time geometry may be responsible for quantum effects,
provided the elementary length $\lambda _{0}$ depends on the quantum
constant $\hbar $ \cite{R91}. However in the theory of elementary particles
the discrimination mechanism is more important, than the quantum effects.

\section{Concluding remarks}

It is impossible to investigate properly microcosm and structure of
elementary particles without a perfect knowledge of geometry. Unfortunately,
we know only axiomatizable geometries, which do not include granular
space-time geometries. The granular geometries generate discrimination
mechanism, which is necessary for explanation and calculation of discrete
characteristics of elementary particles. The axiomatizable geometries cannot
take into account such properties of a geometry as discreteness and limited
divisibility. They cannot generate a discrimination mechanism.

Without a proper knowledge of geometry, we are forced to compensate our
mathematical illiteracy by exotic hypotheses, beginning from quantum
principles and finishing by many-dimensional geometries. Besides, the
investigation strategy, based on finding and correction of mistakes, is a
safe strategy. Correcting mistakes in our knowledge of geometry, I realize
the safe investigation strategy.

To understand interrelation between different kinds of geometries, it is
useful to know interrelation of three different representations of the
proper Euclidean geometry \cite{R2007}

\end{document}